\documentclass{llncs}

\usepackage{natbib}
\usepackage{amsmath}
\usepackage{amssymb}
\usepackage{latexsym}
\usepackage{epsfig}
\usepackage{psfig}
\usepackage{color}
\usepackage{enumerate}

\newtheorem{observation}{Observation}

\newlength{\boxwidth}
\makeatletter
\newcommand{\rset}{\mbox{{\normalfont I\hspace{-.4ex}R}}}

\newcommand{\hide}[1]{}

\renewcommand{\l}{\ell}

\newcommand{\comment}[1]{}

\begin{document}

\title{The Price of Selfish Stackelberg Leadership in a Network Game}

\author{Paul W. Goldberg\inst{1} \and Pattarawit Polpinit\inst{2}}

\institute{University of Liverpool, Department of Computer Science,\\
Ashton Building, Ashton Street Liverpool L69 3BX, U.K.\\
Research supported by the EPSRC grant
``Algorithmics of Network-sharing Games''.\\
\email{P.W.Goldberg@liverpool.ac.uk}
\and
University of Liverpool, Department of Computer Science, \\
Ashton Building, Ashton Street, Liverpool L69 3BX, U.K.\\
\email{Polpinit@liverpool.ac.uk}}

\maketitle

\begin{abstract}
\noindent We study a class of games in which a finite number of agents
each controls a quantity of flow to be routed through a network, and
are able to split their own flow between multiple paths through the
network. Recent work on this model has contrasted the social cost of
Nash equilibria with the best possible social cost.

Here we show that additional costs are incurred in situations where a
selfish ``leader'' agent allocates his flow, and then commits to that
choice so that other agents are compelled to minimise their own cost
based on the first agent's choice. We find that even in simple
networks, the leader can often improve his own cost at the expense of
increased social cost. Focussing on the 2-player case, we give upper
and lower bounds on the worst-case additional cost incurred.
\end{abstract}

\section{Introduction}

Imagine that two firms wish to route traffic from a source to a
destination through a shared network.  Any link suffers from a delay
(also called latency) that increases with the amount of traffic that
it attracts, and both firms want to minimise their own total delay.
It is known (e.g.~\cite{HTW,CCSM}) that the resulting social cost (sum
of individuals' delays) is suboptimal, even for simple networks.  If
for example some but not all links are privately-owned, there is a
tendency for both firms to over-use the shared link, in order to
relieve pressure on the privately-owned links (Catoni and
Pallottino~\cite{CP}, Cominetti et al.~\cite{CCSM}).

Viewing this as a non-cooperative two-player game, suppose now one of
the firms (player 1) is ``forceful'', and the other one (player 2) is
``pliant''. Player~1 may find that it pays to over-use a shared link
even more than before, provided that player~2 responds by moving some
of his own traffic away from the shared link and onto player 2's private
links. As a consequence, it turns out that player~1's total delay may
fall, but player~2's total delay increases by a greater amount, thus
increasing the social cost.

One way to model forceful and pliant players, is to let player 1 have
\emph{Stackelberg leadership}: player~1 selects his strategy, namely
the split of his own flow between the links available to him.  Then,
player 2 chooses his strategy based on player 1's choice, under the
assumption that player 1 will not subsequently change his decision.

Thus there may be a social cost of Stackelberg leadership
over and above the cost of selfish decentralised
decision-making. In this paper we focus on a simple and
well-known setting in which the players have access to a set
of shared ``parallel links''. Note that this is more restrictive
than the scenario described above in that there are no private
links. We give a simple example of how selfish stackelberg leadership
(which we usually abbreviate to SSL)
may nevertheless have a positive cost in this setting, and
motivated by that example, we establish an upper bound on
the price of SSL.

\subsection{Summary of results}

Our focus is on the 2-player atomic-splittable case, with parallel
links having linear latency functions.
In Subsection~\ref{sec:lowerbound} we show that if there exists a
player having Stackelberg leadership, then the social cost may be
higher than in the Nash-Cournot setting. Furthermore, the remaining
flow may even be disadvantaged as a direct result of being controlled
by a single player, rather than a Wardrop flow. This situation arises
in a very simple setting in which two players both have access to just two
links having affine linear latency functions.
This furnishes a lower bound on the price of selfish Stackelberg
leadership, of a multiplicative factor 1.057.

%We show that when some links are
%private, there may be a price of selfish Stackelberg leadership (SSL)
%even with homogeneous linear latencies.

Subsection~\ref{sec:lessThanFourOverThree} gives our main result, a
contrasting upper bound.  We analyse games with two players each
needing to route splittable flow through a shared network of parallel
links having linear latency functions. If the latency functions are
homogenous, there is no cost of SSL. However, for the case of affine
latency functions, we show that the worst-case price of SSL is a
multiplicative constant (thus, independent of the number of links), at
most $1.322$.

\subsection{Related work}

A large body of recent work (initiated mainly by Roughgarden and
Tardos~\cite{RT, Rbook}) has studied from a game-theory perspective,
how selfishness can degrade the overall performance of a system that
has multiple (selfish) users. Much of this work has focused on
situations where users have access to
shared resources, and the cost of using a resource increases as the
resource attracts more usage. Our focus here is on the ``parallel
links'' network topology, also referred to as scheduling jobs
to a set of load-dependent machines, which is one of the most commonly
studied models (e.g. \cite{CV, KLO, KP, MS, ORS, R}). Papers such
as~\cite{AAE, CV, KP} have studied the price of anarchy for these
games in the ``unsplittable flow'' setting, where each user may
only use a single resource. In contrast we study the
``splittable flow'' setting of~\cite{ORS}.
This version (finitely many players, splittable flow) was shown
in~\cite{ORS, Rosen} to possess unique pure Nash equilibria (see
Definition~\ref{def:ne}). Hayrapetyan et al.~\cite{HTW} study the cost
of selfish behaviour in this model, and compare it with the cost
of selfish behaviour in the Wardrop model (i.e. infinitely many
infinitesimal users).

Stackelberg leadership refers to a game-theoretic situation where one
player (the ``leader'') selects his action first, and commits to
it. The other player(s) then choose their own action based on the
choice made by the leader.  Recent work on Stackelberg scheduling
in the context of network flow
(e.g.~\cite{CSM,R,S}), has studied it as a tool to mitigate the
performance degradation due to selfish users. The flow that is
controlled by the leader is routed so as to minimise social cost in
the presence of followers who minimise their own costs.  In contrast,
here we consider what happens when the leading flow is controlled by
another selfish agent.  We show here that the price of
decentralised behaviour goes up even further in the presence of a
Stackelberg leader.

Other papers that consider finitely many players each of which may
split their flow amongst the available paths are~\cite{CK, CCSM, R2}.
Christodoulou and Koutsoupias~\cite{CK} study the price of anarchy in
a similar model to this work, but they consider the social cost as
either the maximum cost of a player or the average of the
players'~costs. Cominetti et al.~\cite{CCSM} study the price of
selfish routing in the context of Nash equilibria in this setting.
They give examples of how the aggregation of flow into a finite
number of competing firms, can introduce inefficiency to the outcome.
They give bounds on how much the total Nash cost can differ from the
socially optimal cost.  It is shown in~\cite{CCSM} that in a network
of parallel links with homogeneous linear latencies, Wardrop
equilibria, Nash equilibria and System optimal flows all coincide.
We show that with affine linear latencies, there is a positive cost
of SSL; also there is a positive cost for simple examples
from~\cite{CCSM, CP} involving private and shared links.

\subsection{Model, Notation and Terminology}

Let $m$ be the number of players, and for $i\in \{1,\ldots,m\}$,
player $i$ has a {\em flow} $f^i\in \rset^+$. (In related literature,
``flow'' is sometimes called ``weight'' or ``demand''.)  A {\em
strategy} of player $i$ is a partition of $f^i$ amongst $n$ resources
(or links, available to carry flow), $f^i = \sum_{j=1}^n f^i_j$, where
$f^i_j$ is the non-negative flow assigned to link $j$ by player $i$.
Given total flow $f$, $f_j$ denotes the portion of $f$ on link $j$,
i.e. $f_j=\sum_i f^i_j$.  The network flow scenario being modelled has
all players with a common source and common destination, connected by
$n$ ``parallel links'' through which the players may route their flow.

For $1\leq j\leq n$ let $\ell_j:\rset^+\longrightarrow\rset^+$
denote link $j$'s latency (or cost) function; this maps the load
on $j$ to the cost of using $j$. Latency functions are positive
and non-decreasing. In this paper, we work with linear latency
functions. Link $j$ has latency function $\ell_j(f_j)=a_jf_j+b_j$,
for non-negative numbers $a_j$ and $b_j$.
That functional form is an {\em affine linear} cost;
a {\em homogeneous} linear cost function takes the form
$\ell_j(f_j)=a_jf_j$.

Suppose that $f$ represents an allocation of flow for each
agent across the available resources. For $1\leq j\leq m$,
$f_j$ denotes the flow on $j$. It is sometimes useful to let $\ell_j(f)$
denote $\ell_j(f_j)$, the cost of using link $j$ in $f$.

{\bf Normalisation.}  We assume that links are numbered in ascending
order of $b_j$, so that for $j<j'$,
$b_j \leq b_{j'}$. We will also assume throughout that $b_1=0$. This
is because flow distributions are unaffected by adding a constant to
all the $b_j$'s (by analogy with the common observation that the Nash
equilibria of a game are unaffected by adding a constant to all
payoffs.) Setting $b_1=0$ maximises ratios between alternative social
costs, which is what we are interested in maximising.

We mainly focus on the 2-player case. We usually assume (by re-scaling
as necessary) that the total flow is one (unless we explicitly state
another quantity). In this case we let $\alpha$ denote
player 1's flow, so that $1-\alpha$ is player 2's flow.

\begin{definition}\label{def:cost}
The {\em cost experienced by a player} is the sum, over all paths used
by that player, of the amount of that player's flow on that path
multiplied by the cost of using that path. (That cost is of course
affected by the other players' choices.)  Thus, the cost experienced
by player $i$ in flow $f$ is given by $\sum_j f^i_j\ell_j(f)$.  The
{\em social cost} is the sum of the individual players' costs.
\end{definition}

Note that the social cost can be expressed as the sum over all links,
of the flow on that link multiplied by the cost of using that link;
$C = \sum_j f_j \ell_j(f_j)$.

\begin{definition}\label{def:ne}
A {\em Nash equilibrium} is a set of strategies
$\{f^i_j~:~1\leq i\leq m, 1\leq j\leq n\}$ such that no player can
reduce his cost (as in Definition~\ref{def:cost}) by changing his
own strategy.
\end{definition}

\begin{definition}\label{def:sl}
Our {\em selfish Stackelberg leadership (SSL)} solution concept is
subgame perfect equilibrium: Assume that player~1 is the leader;
player~1 selects his strategy and players $2\ldots m$ form a
$m-1$-player Nash equilibrium with latency functions that have been
modified to take into account player~1's strategy. It is assumed that
player~1 selects his strategy to minimise his own cost under that
assumption on the other players. In the 2-player case studied here,
player 2 must make an optimal allocation of his own flow based
on latency functions that have been affected by player 1's flow
allocation.
\end{definition}

\begin{definition}
Suppose a finite number $m$ of agents want to route flow through a
network.  The {\em price of selfish Stackelberg leadership} is the
ratio between the social costs of the worst (maximum social cost)
SSL solution that occurs when one of the agents is a leader, and the
unique Nash equilibrium that arises in the standard simultaneous
setting.
\end{definition}

\section{Bounding the Price of Selfish Stackelberg Leadership}
\label{sec:the bound}

This section is organised as follows. In Section~\ref{sec:lowerbound}
we describe the simple example that gives the lower bound of $1.057$
on the multiplicative cost of selfish Stackelberg leadership.  In
Section~\ref{sec:socialFlow} we give some basic results about social
optimal flow in this setting. In Section~\ref{sec:properties} we
describe 2 strategies, one for each player, which have useful upper
bounds on the individual costs they incur. (In particular, since they
guarantee each player an individual cost at most the optimal social
cost, they give an upper bound of 2 on the cost of selfish Stackelberg
leadership.) In Section~\ref{sec:fourOverThreeBound} we show that
in fact they give a stronger upper bound of $\frac{4}{3}$.
Finally, in Section~\ref{sec:lessThanFourOverThree} we show by dint
of a more complicated analysis, and by also considering the {\em aloof
strategy} of Roughgarden~\cite{R}, an upper bound of slightly less
than $\frac{4}{3}$.

\subsection{Lower Bound in Simple Symmetric Network}
\label{sec:lowerbound}

We start our investigation of the price of Stackelberg leadership
by considering an example of the simplest kind of network having
a cost of SSL greater than 1. There are two players; player~1 with
a flow of $\frac{3}{5}$ and player~2 with a flow of $\frac{2}{5}$. 
Players want to transfer flows on a network with two parallel links.
Link~1 has a latency function $\ell_1(f_1)=f_1$ and link~2 has
a latency function $\ell_2(f_2)=\frac{6}{5}$.

For this network, the price of selfish Stackelberg leadership is
$\frac{93}{88}\approx 1.057$. In the appendix
(Section~\ref{sec:computeLowerBound}) we evaluate the resulting solutions
and explain how we optimised the constants involved to show that this
is in fact the largest price of SSL that arises in the 2-link affine
linear cost setting.  (For non-linear latency functions we have obtained
a price of SSL of $1.169$, arising when $\ell_1(f_1)=(f_1)^4$ and
$\ell_2(f_2)=5.67$, and players' flows are $f^1=1$, $f^2 = 0.587$.)

\subsection{Social Optimal Flow}
\label{sec:socialFlow}

In this subsection we note some basic facts about the behaviour of
social optimal flow. In~\cite{CP},
socially optimal flow is called ``system equilibrium'' (SE).  We
let $f_{SE}$ denote the social optimal flow.
The following lemma is a special case of Lemma~4.1(b) of
Roughgarden and Tardos~\cite{RT}.

\begin{lemma}\label{socialOptimalFlow}
(Roughgarden and Tardos~\cite{RT}) For affine latency functions
$\ell_j(f_j)=a_jf_j+b_j$, the single-player (socially optimal) flow has
the property that $2a_j f_j + b_j$ is the same for all links $j$ on which
flow is routed.
\end{lemma}

\begin{proof}
The lemma is a special case of Lemma~4.1(b) of
Roughgarden and Tardos~\cite{RT}. A proof is given in the appendix
since the proof is simpler in this special case.\qed
\end{proof}

\begin{observation}\label{factorOfTwo}
If $2a_jf_j+b_j$ is the same for all $j$, then note that latencies
$\ell_j(f_j) = a_jf_j+b_j$ cannot differ by a factor more than $2$.
\end{observation}

One consequence of the above observation is that all latencies
end up within a factor of 2 of each other (for links that carry
non-zero flow).

\begin{lemma}\label{useful}
For affine linear latencies $\ell_j(f_j)=a_jf_j+b_j$,
suppose links $j$ and $j'$ both carry flow in $f_{SE}$
(recall $f_{SE}$ denote socially optimal flow). Assume
$j<j'$ and $b_j<b_{j'}$. Then $\ell_{j'}(f_{SE}) - \ell_j(f_{SE}) =
\frac{1}{2}(b_{j'}-b_j).$
\end{lemma}

\begin{proof}
If $j$ and $j'$ both carry flow, then we noted above that
$2a_jf_j+b_j$ is the same for all $j$.
\[
\begin{array}{rl}
2(\ell_{j'}(f)-\ell_j(f)) = & (2a_{j'}f_{j'}+2b_{j'})-(2a_jf_j+2b_j)
\\

=  & (2a_{j'}f_{j'}+b_{j'})-(2a_jf_j+b_j)+b_{j'}-b_j = b_{j'}-b_j.
\end{array}
\]\qed
\end{proof}

Note that Lemma~\ref{useful} implies the fact (shown in~\cite{HTW})
that in the homogeneous linear case (where all $b_j$'s are zero) the
latencies are equal in socially optimal flow.

Another consequence is that if a socially optimal flow is routed
through a set of links, then given our assumption that links are
indexed in increasing order of $b_j$, their latencies $\ell_j(f_{SE})$
are sorted in increasing order of $j$ (and for links that are used,
differ by half the latency difference when flow is zero).

\subsection{Properties of First and Second player in Stackelberg
equilibria}
\label{sec:properties}

Recall that we abbreviate Selfish Stackelberg Leadership to SSL, and
$f_{SSL}$ denotes the total flow of two players where the first player
is a leader.  $C_{SE}$ and $C_{SSL}$ denote respectively the social
optimal cost and the social cost in $f_{SSL}$.

The first player decides how much of his flow to be routed on each
link, and commits to that decision. From player~2's
perspective, for the affine linear latency functions, the leader has
essentially increased the latency functions by constants corresponding
to the flow he routed on each link. So now the second player should
find an optimal flow using the new latency function given by
$$\l_j(f^2_j) = a_j.f^2_j + (b_j+a_j.f^1_j)$$

\begin{lemma}\label{SSL2factor}
In $f_{SSL}$, every link used by the last player has a latency at
most twice the minimum latency.
\end{lemma}

\begin{proof}
After player~1 plays, the new latency functions are still
inhomogeneous linear. Apply Observation~\ref{factorOfTwo}.\qed
\end{proof}

\begin{paragraph}{\bf Two Strategies.}
We present two simple strategies for player~1 and player~2
that can guarantee each of them an individual cost at most $C_{SE}$.
(Thus a simple combination of these results indicates that
$C_{SSL} \leq 2C_{SE}$.)
It should be pointed out that these strategies are not necessary optimal
for player 1 and 2.
\end{paragraph}

\begin{paragraph}{Strategy 1:}\label{P1_force_optimum}
The general idea is that player 1 allocates his flow $f^1$ such that
when player 2 minimises his own cost, the combined flow is
socially optimal. Provided player 1 can achieve this, it follows
immediately that player 1's cost is at most $C_{SE}$, which is
the total cost for both players.

Denote $S_{SE}$ and $S_2$ sets of links used by the social optimal
flow and player~2 respectively.  Player~1's strategy to make the total
flow socially optimal is to ensure that the difference between any two
link latencies in $S_{2}$ remains the same after he has played.
Recall that links are indexed in increasing order of constant values,
i.e., for $j<j',b_j\leq b_{j'}$ (where $b_j = \l_j(0)$).  The strategy
is as follows:
\begin{enumerate}
\item Compute the social optimal flow $f_{SE}$.
\item Check whether $S_{SE} - S_{2} \neq \emptyset$ by checking if
    $f^2 \geq \sum_{1\leq i\leq|S_{SE}|}{(b_{|S_{SE}|}-b_i)/(2a_i)}$.
    Player~1 fills up all the links in $S_{SE} - S_{2}$ to the level of
    what should occur in $f_{SE}$ : $f^1_j = (f_j)_{SE}$ for every link
    $j$ in $S_{SE} - S_{2}$.
\item Player~1 splits the remainder of his flow among all the links in
    $S_{2}$ such that every links is increased with the same amount of
    latency: $f^1_j = (f^1 - f^1_{(S_{SE} - S_{2})})/(a_j\sum_{j\in
    S_2}{\frac{1}{a_j}})$ for every link $j \in S_2$.
\end{enumerate}
\end{paragraph}

\begin{paragraph}{Strategy 2 : }\label{P2_find_optimum}
We consider the following strategy for player~2:
\begin{enumerate}
\item Compute the social optimal flow (for total flow of both
        players) $f_{SE}$.
\item In link $j$, if $f^1_j < (f_{SE})_j$, player~2 increases the
        flow on $j$ to the $f_{SE}$ level.
        If $f^1_j < (f_{SE})_{j}, f^2_j = (f_{SE})_{j} - f^1_j$ for every
        link $j$.
\end{enumerate}
With this strategy, if on every link player~1 routes less flow than the
social optimal flow then player~2 just makes up the difference. If there
are some links that player~1 routes more than the social optimal flow
then there is enough room on the remaining links for player~2 to use
those without making their flow more than the social optimal flow.
\end{paragraph}

\subsection{Quick Upper Bound of $\frac{4}{3}$}
\label{sec:fourOverThreeBound}

With those strategies we give an upper bound on player~1's cost and
player~2's cost in the following two Lemmas.

\begin{lemma}\label{P1_bound}
Let $\l_{min}$ be the minimal latency of links under the socially
optimal flow $f_{SE}$.  In the SSL setting with linear cost functions,
player~1's cost is at most
$\min\{2\alpha\l_{min}, C_{SE} - (1-\alpha)\l_{min}\}$,
where $\alpha$ is player 1's flow, out of a total flow of 1.
\end{lemma}

\begin{proof}
Applying Strategy~1, player~1 causes the combined flow to be
socially optimal flow $f_{SE}$.
In that scenario, player~1's cost is $C_{SE}$ minus player~2's cost
which is at least all of player 2's flow $f^2$ multiplied by the
minimal latency $\l_{min}$, i.e.
$(1-\alpha)\l_{min}$. Hence player~1's cost is at most
$C_{SE}-(1-\alpha)\l_{min}$. (It is of course possible that player
1 could do better by not using Strategy 1.)

In addition, player~1's cost cannot be any higher than the cost of
him putting all of his flow of $\alpha$ on the maximal latency links.
Furthermore, from Observation \ref{factorOfTwo}, the maximal latency
is at most $2\l_{min}$ for links that get used. Hence player~1's cost
is at most $2\alpha\l_{min}$.

Combining those two results, we have player~1's cost is at most
$\min\{2\alpha\l_{min}, \\C_{SE} - (1-\alpha)\l_{min}\}$.\qed
\end{proof}

\begin{lemma}\label{P2_bound}
Let $\l_{min}$ be the minimal latency of links under the socially
optimal flow $f_{SE}$.  In the SSL setting with linear cost functions,
player~1's cost is at most
$\min\{C_{SE} - \alpha\l_{min}, 2(1-\alpha)\l_{min}\}$,
where $\alpha$ is player 1's flow, out of a total flow of 1.
\end{lemma}

\begin{proof}
Applying Strategy~2, player~2's cost is at most $C_{SE}$ minus
player~1's cost which is minimised when all player~1's flow is on the
minimal latency links. Hence player~2's cost is at most $C_{SE} -
\alpha \l_{min}$. (It is of course possible that player 2 could do
better by not using Strategy 2.)

Moreover because player~2's cost is maximised when all player~2's flow
is in the maximal latency links and, for latencies of links that get
used, the maximal latency is at most twice the minimal latency,
player~2's cost is at most $(1-\alpha)\l_{max} \leq 2(1-\alpha)\l_{min}$.

Combining those results, we essentially have $C^2 \leq \min\{C_{SE} -
\alpha \l_{min}, 2(1-\alpha)\l_{min}\}$.\qed
\end{proof}

\begin{theorem}\label{fourOverThreeBound}
In a two-player network model with non-decreasing linear latency
functions, the price of SSL is at most 4/3.
\end{theorem}

\begin{proof}
Recall that $\l_{min}$ denotes the minimal latency of a link
in flow $f_{SE}$.
Note that with one unit of flow in total (which we assured by
rescaling as necessary) $C_{SE} \in [\l_{min}, 2\l_{min}]$.
(Using Observation~\ref{factorOfTwo}.)

First consider when $C_{SE} \geq (3/2)\l_{min}$. In this case, from
Lemma~\ref{P1_bound} and \ref{P2_bound},
$\min\{2\alpha\l_{min}, C_{SE} - (1-\alpha)\l_{min}\} =
2\alpha\l_{min}$ and
$\min\{C_{SE} - \alpha\l_{min}, 2(1-\alpha)\l_{min}\} =
2(1-\alpha)\l_{min}$. Thus $C_{SSL}$ is at most
$2\alpha\l_{min} + 2(1-\alpha)\l_{min} = 2\l_{min}$. Hence the ratio
between $C_{SSL}$ and $C_{SE}$ is bounded by
\[
\begin{array}{ccccc}
\frac{C_{SSL}}{C_{SE}} & \leq & \frac{2\l_{min}}
{(3/2)\l_{min}} & = & \frac{4}{3}.
\end{array}
\]

Next suppose $C_{SE} < (3/2)\l_{min}$. Then
$\min\{2\alpha\l_{min}, C_{SE} - (1-\alpha)\l_{min}\} =
C_{SE} - (1-\alpha)\l_{min}$ and
$\min\{C_{SE} - \alpha\l_{min}, 2(1-\alpha)\l_{min}\} =
C_{SE} - \alpha\l_{min}$. Thus $C_{SLL}$ is at most
$C_{SE} - (1-\alpha)\l_{min} + C_{SE} - \alpha\l_{min} =
2C_{SE} -\alpha\l_{min}$.
Hence the the ratio between $C_{SSL}$ and $C_{SE}$ is at most
\[
\begin{array}{ccccccccc}
\frac{C_{SSL}}{C_{SE}} &\leq& \frac{2C_{SE} - \l_{min}}{C_{SE}}
                       &\leq& 2 - \frac{\l_{min}}{(3/2)\l_{min}}
                       & = & \frac{4}{3}.
\end{array}
\]\qed
\end{proof}

Applying this approach we can bound the price of SSL in more
detail, in terms of the ratio between the social optimum cost
and the minimum latency, or player 1's flow $\alpha$.

\begin{lemma}
Let $C_{SE} = \gamma\l_{min}$ where $1 \leq \gamma \leq
2$. Then
\begin{itemize}
\item if $\gamma \geq \frac{3}{2}$ then the price of selfish
      Stackelberg leadership $\leq \frac{2}{\gamma}$;
\item if $\gamma < \frac{3}{2}$ then the price of selfish
      Stackelberg leadership $\leq \frac{2\gamma - 1}{\gamma}$.
\end{itemize}
\label{CseAwayFromThreeHalf}
\end{lemma}

\begin{proof}
When $\gamma \geq 3/2$ then $C_{SSL} \leq 2\l_{min}$. Hence the price of
SSL is upper bounded by
\[
\begin{array}{ccccc}
\frac{2\l_{min}}{C_{SE}} & \leq & \frac{2\l_{min}}{\gamma\l_{min}}
& = & \frac{2}{\gamma}.
\end{array}
\]

Next suppose $\gamma < 3/2$ then $C_{SSL} \leq 2C_{SE} - \l_{min}$.
Hence the price of SSL is upper bounded by
\[
\begin{array}{ccccc}
\frac{2C_{SE} - \l_{min}}{C_{SE}} & = & 2 - \frac{\l_{min}
}{\gamma\l_{min}}
& = & \frac{2\gamma - 1}{\gamma}.
\end{array}
\]\qed
\end{proof}

\begin{lemma}
Let $\alpha = f^1$ where $0 \leq \alpha \leq 1$ then:
\begin{itemize}
\item if $\alpha \geq \frac{1}{2}$ then the price of selfish
         Stackelberg leadership $\leq 1 + \frac{2(1-\alpha)}{3}$.
\item if $\alpha < \frac{1}{2}$ then the price of selfish
         Stackelberg leadership $\leq 1+\frac{2\alpha}{3}$.
\end{itemize}
\label{awayFromOneHalf}
\end{lemma}
\begin{proof}
If $\alpha \geq 1/2$ then $C_{SSL} \leq C_{SE} + (1-\alpha)\l_{min}$
(note from Lemma~\ref{P1_bound} and \ref{P2_bound}, $C^1 \leq C_{SE}
- (1-\alpha)\l_{min}$ and $C^2 \leq 2(1-\alpha)\l_{min}$). It was
shown in the proof of Theorem~\ref{fourOverThreeBound} that the price of
SSL is maximised when $C_{SE} = (3/2)\l_{min}$. Hence the upper bound
of the the price of SSL is given by
\[
\begin{array}{ccccc}
\frac{C_{SE}+(1-\alpha)\l_{min}}{C_{SE}}
           & \leq & 1 + \frac{(1-\alpha)\l_{min}}{(3/2)\l_{min}}
           & = & 1 + \frac{2(1-\alpha)}{3}.
\end{array}
\]

If $\alpha < 1/2$ then $C_{SSL} \leq C_{SE} + \alpha\l_{min}$
(note $C^1 \leq 2\alpha\l_{min}$ and $C^2 \leq C_{SE} -\alpha\l_{min}$).
Hence the price of SSL is upper bounded by
\[
\begin{array}{ccccc}
\frac{C_{SE}+\alpha\l_{min}}{C_{SE}}
           & \leq & 1 + \frac{\alpha\l_{min}}{(3/2)\l_{min}}
           & = & 1 + \frac{2\alpha}{3}.
\end{array}
\]\qed
\end{proof}

\subsection{An Upper Bound of Less than $\frac{4}{3}$}
\label{sec:lessThanFourOverThree}

We prove an upper bound on the price of SSL of less than 4/3. To do
that we improve the upper bound for player~1's cost in
Lemma~\ref{P1_bound} in which player~1's cost is bounded under the
pessimistic assumption that in a situation where he creates the social
optimal flow it is possible for all of player~1's flow to get the
maximal latency and all of player~2's flow to get the minimal latency.

Now let us define the following property that a socially optimal
flow may or may not have, depending on the latency functions of
the links. The case analysis used in what follows is based on
the property, called the {\em diverse latency property}.

\begin{definition}
We say that a socially optimal flow $f_{SE}$ has the {\em diverse
latency property} (DLP) if at least 1/4 of the flow gets a latency
of at most $1.16\l_{min}$ and at least 1/4 of the flow gets a latency
of at least $1.84\l_{min}$.
\end{definition}

We start by proving an upper bound in the case that the DLP is
satisfied, then we use an alternative proof in the case that it is not
satisfied. When the DLP is satisfied, instead of using Strategy 1, we
apply a strategy called the \emph{aloof} strategy~\cite{R} that
corresponds more with what selfish player~1 would do.

\begin{definition} Aloof strategy: (Roughgarden~\cite{R}): player~1
routes $f^1$ optimising his cost in player~2's absence: compute
the socially-optimal flow for a total flow volume of $f^1$.
\end{definition}

\begin{lemma}\label{P1_bound_hold}
If the DLP is satisfied by $f_{SE}$, then player~1's cost is at most
$1.915\alpha\l_{min}$ where $\alpha = f^1$.
\end{lemma}

\begin{proof}
Let $S_{min}$ and $S_{max}$ be the sets of links whose latencies in
$f_{SE}$ are at most $1.16\l_{min}$ and at least $1.84\l_{min}$
respectively. Note that we refer to the latency of $S_{min}$
(respectively, $S_{max}$) to mean the maximal latency of links in
those sets. We will assume in this proof that player~1 uses the aloof
strategy. We exploit the fact that when the DLP holds,
player 1 gets a better performance using the aloof strategy instead
of Strategy 1 described earlier.

Using the DLP assumption and Lemma~\ref{useful}, the difference
between the values $\ell_j(0)$ (i.e. the marginal costs of links when
flow is zero) in $S_{max}$ and $S_{min}$ is at least 2 times the
difference between the minimal latency in $S_{max}$ and the maximal
latency in $S_{min}$, i.e. $2(1.84\l_{min} - 1.16\l_{min}) =
1.36\l_{min}$.

Next we consider $f_{SSL}$. With the aloof strategy, after player~1
has played, the latency difference between $S_{max}$ and $S_{min}$
is at least half the difference between the value $\ell_j(0)$
in $S_{max}$ and
the value $\ell_j(0)$ in $S_{min}$.
Hence the latency difference is at least
$\frac{1}{2}(1.36\l_{min}) = 0.68\l_{min}$.
Thus if player~1 uses $S_{max}$ then the latency of $S_{max}$
minus the latency of $S_{min}$ after player~1 has played is at least
$0.68\l_{min}$. And if player~1 does not use $S_{max}$ then the
difference is higher than $0.68\l_{min}$. Essentially this implies
that in player~2's perspective the fixed cost in $S_{max}$ is at
least $0.68\l_{min}$ more than that in $S_{min}$.
By Lemma~\ref{SSL2factor} and
Observation~\ref{factorOfTwo}, there is not enough flow in total
for the latency in $S_{\max}$ of links used by either player
to be $>2\ell_{\min}$.
Therefore the latency in $S_{min}$ after player~2 has played is at
most $2\l_{min} - \frac{1}{2}(0.68\l_{min}) = 1.66\l_{min}$.

With the DLP assumption, because there is at least $\frac{1}{4}$ of the
total flow in $S_{min}$, player~1 is guaranteed to have at least
$\frac{1}{4}$ of his flow in $S_{min}$. Hence player~1's cost is at most
$\frac{1}{4}\alpha 1.66\l_{min} + \frac{3}{4}\alpha 2\l_{min} =
1.915\alpha\l_{min}$.\qed
\end{proof}

For the case when the DLP is not satisfied, we prove the following upper
bound on player~1's cost.

\begin{lemma}\label{P1_bound_not_hold}
If in $f_{SE}$ at most $\frac{1}{4}$ of the flow gets a latency of at
most $1.16\l_{min}$ or at most $\frac{1}{4}$ of the flow gets a latency
of at least $1.84\l_{min}$ then player~1's cost is at most
$\max\{C_{SE} - \l_{min}/4 - 1.16\l_{min}(\frac{3}{4}-\alpha),
\frac{1}{2}\l_{min} + (\alpha - \frac{1}{4})(1.84\l_{min})\}$.
\end{lemma}

\begin{proof}
If player~1 uses Strategy 1, he can ensure that the combined flow is
socially optimal. In that scenario, player~1's cost is at most the
social optimal cost minus player~2's cost which is lowest when all
player 2's flow $f^2$ gets the latency of $\l_{min}$. However, noting
the first alternative of the given assumption, suppose at most
$\frac{1}{4}$ of the total flow gets latency $\l_{min}$, and the rest
gets at least $1.16\l_{min}$. Hence player~1's cost is at most $C_{SE}
- \l_{min}/4 - 1.16\l_{min}(\frac{3}{4}-\alpha)$.

Alternatively player~1's cost is maximised when all of $f^1$ is in
maximal latency links. However, from the second alternative of the
given assumption, only $\frac{1}{4}$ of the flow gets the latency more
than $1.84\l_{min}$ and the rest of the flow gets the latency of at
most $1.84\l_{min}$. Hence player~1's cost can be at most
$\frac{1}{4}(2\l_{min}) + (\alpha -\frac{1}{4})(1.84\l_{min})
= \frac{1}{2} \l_{min} + (\alpha - \frac{1}{4})(1.84\l_{min})$.\qed
\end{proof}

\begin{theorem}
In a two-player model with non-decreasing linear latency function, the
price of selfish Stackelberg leadership is at most 1.322.
\end{theorem}

\begin{proof}
We identify the scope we should restrict our attention to.
Lemma~\ref{awayFromOneHalf} shows that if $\alpha = f^1$ then the
upper bound improves to $1+\frac{2(1-\alpha)}{3}$
if $\alpha \geq 1/2$ and $1+\frac{2\alpha}{3}$ if $\alpha < 1/2$.
Hence we only consider $\alpha \in [0.483, 0.517]$
since Lemma~\ref{awayFromOneHalf} gives the upper bound of less than
1.322 for $\alpha$ outside this range. Similarly we only consider
$C_{SE}$ in the range $[1.474\l_{min}, 1.513\l_{min}]$ since
Lemma~\ref{CseAwayFromThreeHalf} suggests the price of less than 1.322
outside this range.

  From Lemma~\ref{P2_bound}, player~2's cost is at most $\min\{C_{SE} -
\alpha\l_{min},~2(1-\alpha)\l_{min}\}$. For player~1's cost, we have two
upper bounds, one for when the DLP holds and the other for when the
DLP does not hold.

First, when the DLP is satisfied, we have the upper bound on player~1's
cost of $1.915\alpha\l_{min}$ from Lemma~\ref{P1_bound_hold}. Combining
with player~2's cost, the price of SSL is upper bounded by
\[
\begin{array}{ccl}
\frac{1.915\alpha\l_{min} + \min\{C_{SE} - \alpha\l_{min}, ~2(1-\alpha)\l_{min}\}}{C_{SE}}
    &\leq& \frac{1.915\alpha\l_{min} + C_{SE} - \alpha\l_{min}}{C_{SE}}\\
    & = & 1 + \frac{0.915\alpha\l_{min}}{C_{SE}}\\
    &\leq& 1 + \frac{0.915(0.517)\l_{min}}{1.474\l_{min}}\\
    &<& 1.321
\end{array}
\]

Second, when the DLP is not satisfied, player~1's cost is upper
bounded by $\max\{C_{SE}
-\l_{min}/4-1.16\l_{min}(\frac{3}{4}-\alpha),~ \frac{1}{2}\l_{min} +
(\alpha - \frac{1}{4})(1.84\l_{min})\}$.  Within this proof, we use
(1) and (2) to denote the expressions $C_{SE} -\frac{1}{4}\l_{min}
-1.16\l_{min}(\frac{3}{4}-\alpha)$ and $\frac{1}{2}\l_{min} + (\alpha
- \frac{1}{4})(1.84\l_{min})$ respectively. Thus the upper bound for
player~1's cost can be represented with $\max\{(1), (2)\}$. We prove
the upper bound by exhaustive case analysis as follows:

\begin{enumerate}
\item Suppose $C_{SE} \geq (3/2)\l_{min}$, we consider $\alpha$:
\begin{itemize}
\item Suppose $\alpha < 1/2$: In this scenario $\max\{(1),(2)\} = (1)$.
Hence the price of SSL is upper bounded by:
\[
\begin{array}{ccl}
\frac{(1) + \min\{C_{SE} - \alpha\l_{min},
~2(1-\alpha)\l_{min}\}}{C_{SE}}
    &\leq& \frac{C_{SE} -\l_{min}/4-1.16\l_{min}(3/4-\alpha) +
        2(1-\alpha)\l_{min}}{C_{SE}}\\
    &=& 1 + \frac{(0.88-0.84\alpha)\l_{min}}{C_{SE}}\\
    &\leq& 1 + \frac{(0.88-0.84(0.483))\l_{min}}{(3/2)\l_{min}}\\
    &<& 1.317
\end{array}
\]
\item Suppose $\alpha \geq 1/2$: Now we have player~2's cost $\leq
  2(1-\alpha)\l_{min}$, but player~1's cost is still at most
  $\max\{(1),(2)\}$. The upper bound on the price of SSL can be
  considered in two cases:
\begin{enumerate}
\item
$\max\{(1),(2)\}=(1)$; the price is at most
$\frac{(1) + 2(1-\alpha)\l_{min}}{C_{SE}}$,
\[
\begin{array}{ccl}
\frac{(1) + 2(1-\alpha)\l_{min}}{C_{SE}}
       &=& \frac{C_{SE} -\l_{min}/4-1.16\l_{min}(3/4-\alpha) +
             2(1-\alpha)\l_{min}}{C_{SE}}\\
       &=& 1 + \frac{(0.88 - 0.84\alpha)\l_{min}}{C_{SE}}\\
       &\leq& 1 + \frac{(0.88 -
             0.84(0.5))\l_{min}}{(3/2)\l_{min}}\\
       &<& 1.307
\end{array}
\]
\item
or $\max\{(1),(2)\}=(2)$; the price is at most
$\frac{(2) + 2(1-\alpha)\l_{min}}{C_{SE}}$,
\[
\begin{array}{ccl}
\frac{(2) + 2(1-\alpha)\l_{min}}{C_{SE}}
     &=& \frac{\l_{min}/2 + (\alpha - 1/4)(1.84\l_{min}) +
           2(1-\alpha)\l_{min}}{C_{SE}}\\
     &=& \frac{(2.04 - 0.16\alpha)\l_{min}}{C_{SE}}\\
     &\leq& \frac{(2.04 - 0.16(0.5))\l_{min}}{(3/2)\l_{min}}\\
     &<& 1.307
\end{array}
\]
\end{enumerate}
\end{itemize}
\item Suppose $C_{SE} < (3/2)\l_{min}$, we consider $\alpha$:
\begin{itemize}
\item Suppose $\alpha < 1/2$: In this condition we have player~2's
      cost is at most $C_{SE} - \alpha\l_{min}$. Hence the upper bound
      of the price of SSL can be considered in two cases:
\begin{enumerate}
\item
$\max\{(1),(2)\}=(1)$; the price is at most
$\frac{(1) + C_{SE} - \alpha\l_{min}}{C_{SE}}$,
\[
\begin{array}{ccl}
\frac{(1) + C_{SE} - \alpha\l_{min}}{C_{SE}}
     &=& \frac{C_{SE} -\l_{min}/4-1.16\l_{min}(3/4-\alpha) + C_{SE}
         - \alpha\l_{min}}{C_{SE}}\\
     &=& 2 + \frac{(-1.12 + 0.16\alpha)\l_{min}}{C_{SE}}\\
     &\leq& 2 + \frac{(-1.12 + 0.16(0.5))\l_{min}}{1.5\l_{min}}\\
     &<& 1.307
\end{array}
\]
\item
or $\max\{(1),(2)\}=(2)$; the price is at most
$\frac{(2) + C_{SE} - \alpha\l_{min}}{C_{SE}}$,
\[
\begin{array}{ccl}
\frac{(2) + C_{SE} - \alpha\l_{min}}{C_{SE}}
     &=& \frac{\l_{min}/2 + (\alpha - 1/4)(1.84\l_{min}) + C_{SE} -
         \alpha\l_{min}}{C_{SE}}\\
     &=& 1 + \frac{(0.04 + 0.84\alpha)\l_{min}}{C_{SE}}\\
     &\leq& 1 + \frac{(0.04 + 0.84(0.5))\l_{min}}{1.474\l_{min}}\\
     &<& 1.313
\end{array}
\]
\end{enumerate}
\item Suppose $\alpha \geq 1/2$: In this condition
$\max\{(1), (2)\} = (2)$. Hence the price of SSL is upper bounded by:
\[
\begin{array}{ccl}
\frac{(2) + \min\{C_{SE} - \alpha\l_{min},
         ~2(1-\alpha)\l_{min}\}}{C_{SE}}
     &\leq& \frac{(2) + C_{SE} - \alpha\l_{min}}{C_{SE}}\\
     &=& \frac{\l_{min}/2 + (\alpha - 1/4)(1.84\l_{min}) +
         C_{SE} - \alpha\l_{min}}{C_{SE}}\\
     &=& 1+\frac{(0.025 + 0.9\alpha)\l_{min}}{C_{SE}}\\
     &\leq& 1+\frac{(0.04 + 0.84(0.517))\l_{min}}{1.474\l_{min}}\\
     &<& 1.322
\end{array}
\]
\end{itemize}
\end{enumerate}

Therefore, when the DLP is not satisfied, we have shown that the
price of SSL is less than 1.322.\qed
\end{proof}

\section{Conclusions}

We have shown that the worst-case price of selfish Stackelberg
leadership is a multiplicative constant that is independent of the
number of links in a parallel-links network. Furthermore we have
identified quite a narrow range for that constant, namely $[1.057,
1.322]$. For non-linear latency functions, we have slightly larger
lower bounds on the price of SSL, as noted in
Section~\ref{sec:lowerbound}. The upper bound of 2 that we noted in
Section~\ref{sec:properties} seems to apply in this case, provided
that we have a network of shared parallel links. It is possible that
for parallel links, the worst-case arises for a 2-link network (by
analogy to~\cite{R3}). Perhaps the main question to ask is whether
there is a more dramatic cost (perhaps depending on the size of the
network) in the setting of more general networks.

One alternative line of work is investigating the price of
SSL in a model of one selfish splittable leader and the rest of the
players each with a negligible fraction of the flow (a Wardrop
flow). We believe that there is no price of SSL in this setting since
the SSL solution is essentially the same as Nash equilibrium.

\bibliographystyle{abbrv}

\begin{thebibliography}{99}

\bibitem{AAE} B.~Awerbuch, Y.~Azar and A.~Epstein.
The Price of Routing Unsplittable Flow.
{\em In STOC}, pp.~57-66 (2005).

\bibitem{CP} S.~Catoni and S.~Pallottino.
Traffic Equilibrium Paradoxes.
{\em Transportation Science}, Vol.~35, No.~2, pp.~240-244,
August 1991.

\bibitem{CK} G.~Christodoulou and E.~Koutsoupias.
The price of anarchy of finite congestion games.
\emph{Proceedings of 37th STOC}, pp.~67-73, May 2005.

\bibitem{CCSM} R.~Cominetti, J.R.~Correa and N.E.~Stier-Moses.
Network Games with Atomic Players.
In {\em ICALP}, (2006).

\bibitem{CSM} J.R.~Correa and N.E.~Stier-Moses.
Stackelberg Routing in Atomic Network Games. (February 2007).
Columbia Working Paper No. DRO-2007-03
Available at SSRN: http://ssrn.com/abstract=987115

\bibitem{CV} A.~Czumaj and B.~V\"ocking.
Tight bounds for worst-case equilibria.
\emph{Proceedings of 13th SODA}, pp.~413-420, January, 2002.

\bibitem{FPT} A.~Fabrikant, C.H.~Papadimitriou and K.~Talwar.
The Complexity of Pure Nash Equilibria.
{\em Proceedings of 36th STOC} pp.~604-612, Jun 2004.

\bibitem{FRV} S.~Fischer, H.~R\"acke and B.~V\"ocking.
Fast Convergence to Wardrop Equilibria by Adaptive Sampling
Methods.
In {\em 38th STOC}, (May 2006).

\bibitem{GZ} I.~Gilboa and E.~Zemel.
Nash and correlated equilibria: Some complexity considerations.
{\em Games and Economic Behavior}, (1989).

\bibitem{HTW} A.~Hayrapetyan, E.~Tardos and T.~Wexler.
The Effect of Collusion in Congestion Games.
{\em Proceedings of 38th STOC} pp.~89-98, May 2006.

\bibitem{KLO} Y.~Korilis and A.~Lazar and A.~Orda.
Capacity Allocation under Noncooperative Routing.
\emph{IEEE Trans.~on Automatic Control} (42), pp.~309--325,
March 1997.

\bibitem{KP} E.~Koutsoupias and C.H.~Papadimitriou.
Worst-Case Equilibria.
In {\em Procs. 16th STACS}, (1999), LNCS Vol.~1563 pp.~404-413.

\bibitem{MS} M.~Mavronicolas and P.~Spirakis.
The price of selfish routing.
\emph{In Proceedings of 33rd STOC}, pp.~510-519, Hersonissos,
Greece, (2001).

\bibitem{ORS} A.~Orda, R.~Rom and N.~Shimkin.
Competitive Routing in Multiuser Communication Networks.
{\em IEEE/ACM Trans.~on Networking} 510-521 (1993).

\bibitem{Rosen} J.B.~Rosen.
Existence and Uniqueness of Equilibrium Points for Concave
N-Person Games.
{\em Econometrica}, 520-534 (1965).

\bibitem{R2} T.~Roughgarden.
Selfish Routing with Atomic Players.
In {\em SODA 16}, pp.~1184-1185 (2005).

\bibitem{R3} T.~Roughgarden.
The Price of Anarchy is Indepedent of the Network Topology.
{\em J.~Comput.~Syst.~Sci} 67(2) 341-364 (2003).

\bibitem{R} T.~Roughgarden.
Stackelberg Scheduling Strategies.
{\em SIAM Journal on Computing} 33(2), pp.~332-350 (2004).

\bibitem{Rbook} T.~Roughgarden.
{\em Selfish Routing and the Price of Anarchy}.
MIT Press (2005).

\bibitem{RT} T.~Roughgarden and \'E. Tardos.
How Bad is Selfish Routing?.
{\em J. ACM} 49(2), pp.~236-259 (2002).

\bibitem{S} C.~Swamy.
The Effectiveness of Stackelberg Strategies and Tolls for
Network Congestion Games.
{\em Procs.~of SODA}, pp.~1133-1142 (2007).

\end{thebibliography}

%\newpage
\section{Appendix}

\subsection{Selfish Stackelberg Leadership in an Asymmetric
Network}
\label{sec:asymmetric}
We consider an example studied in the literature that show a
non-trivial cost of selfish Stackelberg leadership. The network
is asymmetric (having private links).

\begin{figure}[ht]
%\begin{figure}[h]
\centering \epsfig{file=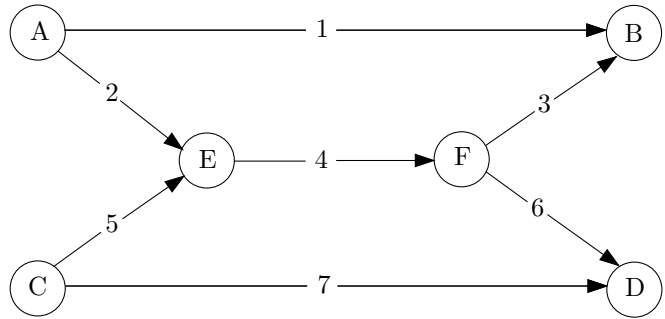, scale = 1}
\caption{Asymmetric network}
\label{fig1}
\end{figure}

\begin{example}
Consider the network depicted in~Figure~\ref{fig1},
studied in~\cite{CP}. There are two origin/destination pairs,
$(A,B)$ and $(C,D)$. $f_j$ is the flow on edge $j$.
The (affine linear) cost functions considered in~\cite{CP}
associated with the links are:
\[
\ell_1(f_1) = f_1 + 30;~~~\ell_4(f_4) = f_4 + 60;
~~~\ell_7(f_7) = f_7;~~~\ell_2 = \ell_3 = \ell_5 = \ell_6 = 0.
\]

Each player has a flow volume of 630.
\end{example}

For this network, \cite{CP} give the following results (an
``atomic'' player is a player with splittable flow that tries to
minimise the overall cost of that flow):

\begin{enumerate}
\item Socially optimum flow (System Equilibrium(SE)):
social cost is 566550.
\item Wardrop equilibrium with infinitely
many infinitesimal users (in~\cite{CP}, the ``user equilibrium'' (UE)):
Social cost is 567000.
\item Nash equilibrium ($(A,B)$-player is atomic; $(C,D)$-player is
Wardrop): social cost is 572400.
\item Nash equilibrium (both players are atomic):
social cost is 576404.
\end{enumerate}

\begin{observation}
For this example from~\cite{CP}, the cost of SSL is greater than 1.
If the $(A,B)$-player is atomic and a selfish
Stackelberg leader, the social cost is 580032 if the $(C,D)$-player
is atomic. If the $(C,D)$-player is a Wardrop flow, the social cost
turns out to be 583538 --- the leader is able to force even more of
the remaining flow onto link 7.
\end{observation}

Cominetti et al.~\cite{CCSM} show that for this network topology there
is a price of anarchy for {\em homogeneous} latency functions.  They
consider an example where the $(A,B)$ and $(C,D)$ players have flows
of 2 and 3 respectively, and edges have latencies $\ell_1(f_1)=f_1$,
$\ell_4(f_4)=f_4$, $\ell_7(f_7)=2f_7$. The optimal (balanced) flow has
cost 10 and the Nash equilibrium has cost approximately $10.47$.

\begin{observation}
In this example of~\cite{CCSM}, the price of SSL is
$10.52/10.47 \approx 1.0048$.

If the $(A,B)$-player has Stackelberg leadership then the cost is
$10.52$ (his flow on link 1 is $\frac{16}{11}$, flow of follower
on link 7 is $\frac{12}{11}$) and if the $(C,D)$-player has
Stackelberg leadership the cost is $10.50$ (flow of
$(A,B)$-player on link 1 is $\frac{3}{2}$, of $(C,D)$-player on link 7
is $1$).
\end{observation}

It is sometimes better for follower's flow to be controlled by
continuum non-atomic players rather than an atomic player as we
demonstrate in the following example.

\begin{example}
Consider a 2-node, 2-parallel-link network, in which the
first link has latency function $\l_1(f_1) = f_1$ and the second link
has latency function $\l_2(f_2) = f_2 + 1$. Let us suppose that the 
leader has flow $f^1 = \frac{1}{2}$ unit and the follower has flow
$f^2 = \frac{1}{2}$ unit to be routed. Then if the follower is 
splittable player, the Stackelberg equilibrium is
$\left(\{f^1_1, f^1_2\},\{f^2_1, f^2_2\}\right) =
\left(\{\frac{1}{2},0\},\{\frac{3}{8},\frac{1}{8}\}\right)$ as a 
result of that the leader gets the cost of
$(\frac{1}{2})(\frac{1}{2}+\frac{3}{8}) = \frac{7}{16} \approx 0.44$;
and the follower gets the cost of
$(\frac{3}{8})(\frac{1}{2}+\frac{3}{8}) +
(\frac{1}{8})(\frac{1}{8}+1) = \frac{15}{32} \approx 0.47$.

If the follower is a collection of infinitely many players each 
controlling a negligible fraction of the flow, the Stackelberg solution 
is $\left(\{\frac{3}{8},\frac{1}{8}\}, \{\frac{1}{2},0\}\right)$
which incurs the leader's cost of $0.47$, and the follower's cost 
of $0.44$.
\end{example}

\subsection{Computing the lower bound for 2 symmetric links}
\label{sec:computeLowerBound}

\begin{paragraph}{Standard Nash equilibrium.}
Players 1 and 2 solve the problems $\min C^1$ and $\min C^2$
accordingly, where $C^i$ is the cost to player $i$. In this
proof we assume player~1's flow is 1 and player~2's flow is $r$.
Player~1 solves an optimisation problem $\min C^1$.
\end{paragraph}

\begin{eqnarray}\label{C1expression}
C^1 &=& (a_1(f^1_1 + f^2_1) + b_1) f^1_1 + (a_2(f^1_2 + f^2_2)
        + b_2) f^1_2\nonumber\\
    &=& (a_1(f^1_1 + f^2_1) + b_1) f^1_1 + (a_2(1 - f^1_1 + r - f^2_1)
        + b_2) (1-f^1_1)
\end{eqnarray}
where $f^1_2$ and $f^2_2$ are substituted with $1-f^1_1$ and
$r-f^2_1$ respectively.

Of course, $C^2$ has a similar expression. By setting the derivatives
to zero and solving algebraically, we get the following expression for
the social cost:

\begin{eqnarray}\label{NEcost}
C^1 + C^2 &=& \frac{9(1+r)( a_2b_1 + a_1(a_2+b_2+a_2r)) - 2(b_1
        - b_2)^2}{9(a_1+a_2)}
\end{eqnarray}

\bigskip
\begin{paragraph}{Cost with Selfish Stackelberg Leadership (SSL).}
Player~1 selects his action in the game first. He predicts what
player~2 will do by solving $\min C^2$ in terms of his own flows, and 
chooses his own flows to minimise his own cost under that assumption. 
Again, this can be solved algebraically by finding expressions for 
$f^2_1$ and $f^2_2$ in terms of $f^1_1$ and $f^1_2$, plugging these
expressions into~(\ref{C1expression}), and minimizing over $f^1_1$ and
$f^1_2$. The social cost with SSL is given by
\begin{equation}\label{SSLcost}
C^1 + C^2 = \frac{16 (1+r) (a_2b_1 + a_1(a_2 + b_2 + a_2r)) -
3(b_1 - b_2)^2}{16(a_1 + a_2)}
\end{equation}
\end{paragraph}

\begin{paragraph}{The price of SSL:}
This is the ratio of~(\ref{SSLcost}) and~(\ref{NEcost})
which is given by
\begin{equation}\label{POS-with-dif-weight-players}
  \frac{9( 16(1+r) ( a_2b_1 + a_1(a_2 + b_2 + a_2r) ) - 3(b_1 -
    b_2)^2)}  {16 ( 9(1+r) ( a_2b_1 + a_1 (a_2 + b_2 + a_2r)) -
    2(b_1 - b_2)^2)}
\end{equation}
\end{paragraph}

We have to make sure that all flows are feasible. In other words
$0 \leq f^1_1$, $f^1_2 \leq 0$ and $0 \leq f^2_1$,$f^2_2 \leq r$.
Therefore we have
\begin{eqnarray}
2a_1 + b_1 - b_2 \geq 0 \nonumber\\
2a_2 - b_1 + b_2 \geq 0 \nonumber\\
3a_1r + b_1 - b_2 \geq 0 \nonumber\\
3a_2r - b_1 + b_2 \geq 0 \nonumber
\end{eqnarray}

We maximise the Price of SSL in
Equation~(\ref{POS-with-dif-weight-players}) under the above
constraints.  The maximum is $93/88 \approx 1.057$, achieved
when $2a_1 = b_2$, $a_2 = b_1 = 0$ and $r = 2/3$.

\begin{paragraph}{The solutions:}
The standard Nash solution has $f^1_1=\frac{2}{3}$;
$f^2_1=\frac{2}{3}$. The SSL solution has $f^1_1=1$;
$f^2_1=\frac{1}{2}$; player~1 (the leader) has forced player~2 to
displace some of his flow onto link 2. If player~2 is a Wardrop flow,
solutions to both versions have  $f^1_1=\frac{2}{3}$;
$f^2_1=\frac{2}{3}$ (the same as the standard Nash solution) ---
thus player~2 is better off as a Wardrop flow, than as a
``coalition'' of the infinitesimal users that constitute a Wardrop
flow.
\end{paragraph}

\begin{observation}
There is no cost to Stackelberg leadership in the special case of
symmetric access to parallel links having homogeneous linear costs.
Furthermore the Nash equilibrium is the same as in the standard
Cournot setting.
\end{observation}

\subsection{Proof of Lemma~\ref{socialOptimalFlow}}

\begin{proof}
For $1\leq j\leq n$ let $f_j$ denote the flow in link $j$.
The benefit of transferring $\epsilon$ from $j$ to $j'$ is the
new cost minus the old cost, so it is given by
\[
\begin{array}{rl}
  & (f_j - \epsilon). \ell_j (f_j - \epsilon) +
(f_{j'} + \epsilon). \ell_{j'} (f_{j'} + \epsilon)
 - [
f_j.\ell_j(f_j) + f_{j'} \ell_{j'}(f_{j'})
]
\\

= & f_j. \Bigl(\ell_j   (f_j    - \epsilon) - \ell_j   (f_j)\Bigr) +
f_{j'}.\Bigl(\ell_{j'}(f_{j'}   + \epsilon) -
\ell_{j'}(f_{j'})\Bigr) - \epsilon.\ell_j(f_j - \epsilon) +
\epsilon.\ell_{j'}(f_{j'}+\epsilon).
\end{array}
\]

In the limit of $\epsilon \longrightarrow 0$ this is equal to

\[
- \epsilon.f_j.   \frac{\partial(\ell_j   (f_j))}{\partial f_j}
+ \epsilon.f_{j'}.\frac{\partial(\ell_{j'}(f_{j'}))}{\partial f_{j'}}
- \epsilon.\ell_j(f_j - \epsilon) + \epsilon.\ell_{j'}(f_{j'}+\epsilon)
\]

Set the above to zero for optimality; divide by $\epsilon$,
note that $\ell_j(f_j+\epsilon) \longrightarrow \ell_j(f_j)$, hence

\[
f_j  \frac{\partial(\ell_j(f_j))}{\partial f_j}
 + \ell_j(f_j)
=
f_{j'} \frac{\partial(\ell_{j'}(f_{j'}))}{\partial f_{j'}}
 + \ell_{j'}(f_{j'}).
\]
In the linear context where $\ell_j(f_j) = a_j f_j + b_j$, we
are saying that for all $j$, $f_j.a_j + a_j f_j + b_j$ is the
same, i.e. the result follows.\qed
\end{proof}

\end{document}